\documentclass[sigconf]{acmart}

\renewcommand\footnotetextcopyrightpermission[1]{}
\setcopyright{none}

\acmConference[SBCARS 2026]{20th Brazilian Symposium on Software Components, Architectures, and Reuse}{September 8--11, 2026}{S\~ao Paulo, SP, Brazil}

\AtBeginDocument{
    
    \hypersetup{hidelinks}
}

\usepackage{booktabs}
\usepackage{seqsplit}
\usepackage{xurl}

\newcommand{\modelid}{\texttt{\seqsplit{gpt-5.4-2026-03-05}}}
\newcommand{\isoqm}{ISO/IEC~25010:\allowbreak 2023}
\newcommand{\artifacturl}{https://doi.org/10.5281/zenodo.21880022}
\newcommand{\rqthreestdev}{\texttt{results/}\allowbreak\mbox{\texttt{rq3\_per\_problem\_stdev.json}}}
\newcommand{\wonestabilitydir}{\texttt{results/}\allowbreak\mbox{\texttt{w1-stability}}}
\newcommand{\resultsnumbers}{\texttt{\seqsplit{results\_numbers.json}}}

\begin{document}

\title{Does ISO-Grounded NFR Specification Improve LLM Code Generation? A Comparison of Rich and Structured Interventions against a Natural-Language Baseline}

\renewcommand{\shorttitle}{ISO-Grounded NFR Enrichment vs.\ NL-Simple Baseline}

\renewcommand{\shortauthors}{Pereira and Garcia}

\author{Jo\~{a}o Pedro Monteiro Pereira}
\affiliation{%
  \institution{Centro de Inform\'{a}tica, Universidade Federal de Pernambuco}
  \city{Recife}
  \state{PE}
  \country{Brazil}}
\email{jpmp2@cin.ufpe.br}

\author{Vinicius Cardoso Garcia}
\affiliation{%
  \institution{Centro de Inform\'{a}tica, Universidade Federal de Pernambuco}
  \city{Recife}
  \state{PE}
  \country{Brazil}}
\email{vcg@cin.ufpe.br}

\begin{abstract}
\begingroup\sloppy
In LLM-based code generation, Non-Functional Requirements (NFRs) are often specified as terse one-line phrases. We ask whether grounding those specifications in \isoqm, either as rich natural-language prose (NL-rich) or as structured JSON (Structured), improves code generated on HumanEval/HumanEval-ET compared to a RobuNFR-style one-line baseline (NL-simple). We evaluate four NFRs (performance, error handling, code smell, readability) with ten prompt variations per condition under a fixed model snapshot and paired non-parametric analysis. \textbf{Primary finding:} ISO-grounded enrichment improves static quality proxies (unreadability density falls across all four NFRs (e.g., Performance $0.88{\rightarrow}0.69$ for NL-rich)) and reduces sensitivity to prompt wording, but does not reliably improve functional correctness; for error handling, extended-test pass rate decreases, suggesting tension between defensive coding patterns and exact-output benchmarks. \textbf{Secondary finding:} when ISO content is held constant, NL-rich and Structured differ negligibly in correctness ($|\delta|{\le}0.023$), indicating that semantic content matters more than JSON-vs-prose format. Practitioners should invest in standard-grounded NFR content rather than serialization form. A fully traceable replication package is provided.
\endgroup
\end{abstract}

\keywords{Non-Functional Requirements, Large Language Models, Code Generation, Prompt Engineering, Software Quality, ISO/IEC 25010, Empirical Software Engineering}

\frenchspacing
\maketitle

\section{Introduction}
Large Language Models (LLMs) are now routinely used to generate source code from natural-language descriptions~\cite{chen2021codex,austin2021mbpp}. While most benchmarks focus on \emph{functional} correctness, real software must also satisfy \emph{Non-Functional Requirements} (NFRs) such as performance efficiency, reliability, and maintainability~\cite{han2024archcode,lai2025nfqc}. Functional correctness is commonly summarized as \textbf{Pass@1}: the fraction of benchmark tasks solved correctly on the first generation attempt. Recent work evaluates how well LLMs honor NFRs and how \emph{robust} outputs are when the same requirement is re-worded. RobuNFR~\cite{lin2025robunfr} shows that adding an NFR to a prompt tends to reduce Pass@1 and increases standard deviation across surface variations of the same requirement.

\textbf{Problem.} In practice, NFRs are often expressed as short, ambiguous natural-language phrases. RobuNFR adopts this style (one line per NFR). Teams that care about software quality may instead document NFRs using product-quality models such as ISO/IEC~25010:2023, either as rich prose or as structured artifacts consumed by architecture and MDE tooling. It remains unclear whether such \emph{enrichment} helps LLMs produce better code, or whether only the \emph{format} (prose vs.\ JSON) matters.

\textbf{Gap.} Prior work establishes that NFRs affect generation and that prompt variation matters~\cite{lin2025robunfr}, and that ISO-grounded quality characteristics are relevant to LLM-generated code~\cite{lai2025nfqc}. What is missing is a controlled comparison of \emph{enrichment level} (NL-simple vs.\ ISO-grounded content) and, secondarily, of \emph{representation form} (NL-rich vs.\ Structured) under fixed ISO content.

\textbf{Investigation.} We evaluate two ISO/IEC~25010:2023-grounded interventions against a RobuNFR-style \textbf{NL-simple baseline} (one-line NFR per task): (i)~\textbf{NL-rich}, a detailed natural-language paragraph; and (ii)~\textbf{Structured}, the same ISO content serialized as JSON. All three conditions use \modelid; the NL-simple baseline was collected in an earlier batch (Apr--May 2026) and was not re-run when interventions executed (June 2026). Ten prompt variations per condition, identical functional task clauses, and HumanEval/HumanEval-ET yield paired per-problem statistics for four NFRs: performance, error handling, code smell, and readability.

We investigate four research questions, ordered by priority:
\begin{itemize}
  \item \textbf{RQ1 (Correctness vs.\ baseline).} Do NL-rich or Structured improve functional correctness (Pass@1, ET-Pass@1) over NL-simple?
  \item \textbf{RQ2 (NFR quality vs.\ baseline).} Do they improve NFR-specific code quality (exception, code-smell, and unreadability densities per ten LOC)?
  \item \textbf{RQ3 (Robustness vs.\ baseline).} Do they reduce sensitivity across ten prompt variations (STDEV) relative to NL-simple?
  \item \textbf{RQ4 (Form; secondary).} When ISO content is held constant, does representation form (NL-rich vs.\ Structured) affect correctness, quality, or robustness?
\end{itemize}

\textbf{Preview of answer.} Enrichment improves \emph{quality proxies} and \emph{robustness} relative to NL-simple, but not functional correctness (and may harm Error Handling ET-Pass@1). When ISO content is fixed, prose and JSON behave similarly. The scientific story is therefore: \emph{what} you specify (ISO-grounded enrichment) matters more than \emph{how} you serialize it (NL-rich vs.\ JSON).

Our contributions are: (i)~paired statistical comparisons of both interventions against NL-simple (primary); (ii)~a content-controlled form comparison (secondary); (iii)~evidence that enrichment narrows prompt sensitivity despite shorter, less lexically diverse prompts; and (iv)~a transparent replication package with traceability from every reported number to its source file.

The remainder of the paper is organized as follows. Section~2 summarizes background; Section~3 positions related work; Section~4 describes the method; Section~5 presents results; Section~6 discusses implications; Section~7 lists threats to validity; Section~8 concludes.

\section{Background}
\subsection{Non-Functional Requirements and ISO/IEC 25010}
NFRs constrain \emph{how} a system behaves rather than \emph{what} it computes. ISO/IEC~25010:2023~\cite{iso25010} defines a product-quality model with characteristics including \emph{Performance Efficiency} (e.g., time behavior, resource utilization), \emph{Reliability} (e.g., fault tolerance), and \emph{Maintainability} (e.g., analysability, modifiability). ISO/IEC~25002:2024~\cite{iso25002} provides the conceptual overview and guidance on how such quality models are used. We use these standards as an external, citable source of NFR \emph{content}, so that NL-rich and Structured encode the same standard-grounded information and differ only in representation form. This design lets us separate \emph{enrichment} (adding ISO-grounded constraints and acceptance criteria beyond a one-line phrase) from \emph{format} (prose vs.\ JSON).

\subsection{LLM Code Generation and Its Evaluation}
Functional correctness of generated code is commonly measured with Pass@1 on HumanEval~\cite{chen2021codex} and MBPP~\cite{austin2021mbpp}. Because the original test suites can be weak, EvalPlus~\cite{liu2023evalplus} augments them with additional inputs (HumanEval+). We additionally report ET-Pass@1 on HumanEval-ET extended oracles, following the RobuNFR evaluation protocol~\cite{lin2025robunfr}, to reduce false positives. We report both Pass@1 and ET-Pass@1. Functional metrics alone cannot establish NFR satisfaction; we therefore complement them with NFR-aligned static-analysis densities (next subsection) and prompt-variation robustness (Section~\ref{sec:results}).

\subsection{Code-Quality Metrics as Densities}
Following RobuNFR~\cite{lin2025robunfr}, NFR-related quality is measured with static analysis normalized by code size: \emph{code-smell density} from Pylint~\cite{pylint} \emph{Refactor} messages and \emph{unreadability density} from Pylint \emph{Convention} messages, both per ten lines of code (LOC); \emph{exception density} as exception-handling statements per ten LOC; and execution time as the mean over repeated runs. Normalizing per LOC is essential because richer or structured prompts can produce longer code, and raw issue counts would otherwise confound size with quality. LOC is computed by the evaluation pipeline.

We selected these proxies because they (i)~are automatable at scale across 164$\times$10 completions per condition, (ii)~map reasonably to Maintainability and Reliability wording in our ISO objects, and (iii)~remain comparable to RobuNFR's reporting style. They do not measure all ISO/IEC~25010 characteristics (e.g., security, compatibility); our four NFRs focus on performance efficiency, fault tolerance, and maintainability-related concerns that admit static or lightweight dynamic signals on HumanEval-sized programs.

\section{Related Work}
\textbf{Robustness of NFR-aware generation.} RobuNFR~\cite{lin2025robunfr} evaluates LLM robustness across four NFR dimensions (design, readability, reliability, performance) using prompt variation, regression testing, and diverse workflows, and reports that including NFRs reduces Pass@1 (by up to 39\%) while increasing standard deviation across re-wordings. RobuNFR's NFR prompts are intentionally terse (one line), mirroring how practitioners sometimes specify quality in ad hoc natural language. Our study extends this methodology with ISO-grounded \emph{enrichment} as the primary factor and treats representation \emph{form} as secondary.

\textbf{Requirement-aware generation.} ArchCode~\cite{han2024archcode} organizes functional and non-functional requirements from textual descriptions via in-context learning and introduces HumanEval-NFR, the first benchmark to evaluate NFRs alongside functional requirements. ArchCode focuses on \emph{improving} requirement satisfaction through retrieval and organization. We instead ask a complementary question: given fixed functional tasks, does \emph{how much} and \emph{how} NFR content is specified change outcomes when the model and benchmark are held constant?

\textbf{Benchmark and protocol choice (RobuNFR vs.\ HumanEval-NFR).} We use HumanEval/HumanEval-ET with RobuNFR's ten-variation protocol rather than HumanEval-NFR for three reasons: (i)~our primary factor is ISO-grounded \emph{enrichment} against a one-line baseline and a secondary \emph{form} contrast with content held constant (not requirement classification or retrieval from unstructured text, which HumanEval-NFR targets); (ii)~RobuNFR's terse NFR prompts align with our NL-simple baseline and document sensitivity under the same functional scaffold; (iii)~HumanEval-ET supplies extended functional oracles across conditions without changing the task distribution. We cite RobuNFR as a \emph{methodological reference} and collect our own NL-simple baseline under \modelid; replication on HumanEval-NFR is future work.

\textbf{Quality characteristics of generated code.} A recent study on quality assurance of LLM-generated code organizes the analysis around ISO/IEC~25010 quality characteristics and combines a literature review, practitioner workshops, and an empirical study~\cite{lai2025nfqc}. This corroborates the relevance of standard-grounded NFRs but does not compare enrichment level against a one-line baseline, nor does it hold ISO content constant while varying form.

\textbf{Structured vs.\ natural-language requirements.} In component-based and model-driven workflows, NFRs often appear as structured attributes (e.g., JSON, DSLs, architecture decision records) linked to quality models. A natural hypothesis is that LLMs might prefer structured input. Our RQ4 directly tests this under content control; the empirical answer is that format alone does not move correctness once ISO content is fixed.

\textbf{NFR specification quality and ambiguity.} Requirements-engineering research consistently finds that NFRs are hard to state precisely and are frequently left ambiguous. A controlled experiment shows that NL requirements-quality defects (notably ambiguous pronouns) propagate into downstream activities such as domain modeling~\cite{frattini2024reqquality}; a survey shows that NFRs in particular tend to be specified late, in uncertain and unstable terms, and revised throughout a project~\cite{viviani2023nfrinstability}. Almonte et al.~\cite{almonte2025nfrllm} use LLMs to derive NFRs grounded in \isoqm\ from functional requirements and report strong agreement with expert assessments. Together these works establish (i)~that underspecified requirement phrasing, including for NFRs, is a real and costly problem and (ii)~that ISO/IEC~25010 is a natural backbone for making NFRs explicit. We differ in goal and design: rather than \emph{generating} or \emph{classifying} NFRs, we treat ISO-grounded content as a controlled \emph{prompt-input manipulation} and measure its effect on the generated code (correctness, static-quality densities, and robustness) against a terse NL-simple baseline, directly connecting the well-documented difficulty of operationalizing NFRs in measurable terms to a concrete generation outcome (Section~\ref{sec:disc-benchmark-limits}).

\textbf{Research gap.} Prior work shows that (i)~NFRs affect LLM outputs, (ii)~prompt variation matters, and (iii)~ISO quality models are relevant. Missing is a paired, per-problem comparison of \emph{enrichment} (NL-simple vs.\ ISO-grounded interventions) with a secondary, content-controlled form contrast. Our primary contribution fills the first gap; RQ4 fills the second.

\textbf{Evaluation infrastructure.} HumanEval~\cite{chen2021codex}, MBPP~\cite{austin2021mbpp}, and EvalPlus~\cite{liu2023evalplus} provide functional-correctness scaffolding; HumanEval-ET extended oracles follow RobuNFR~\cite{lin2025robunfr}. Pylint~\cite{pylint} supplies the static-analysis metrics we reuse.

\section{Research Method}
\label{sec:method}
\subsection{Overview and Experimental Conditions}
We compare two \textbf{interventions} against a \textbf{NL-simple baseline}:
\begin{itemize}
  \item \textbf{NL-simple (baseline)}: the RobuNFR-style one-line NFR phrase per task (collected in this study, same model snapshot).
  \item \textbf{NL-rich (intervention)}: ISO/IEC~25010 content written as a rich natural-language paragraph.
  \item \textbf{Structured (intervention)}: the \emph{same} ISO content serialized as JSON (attribute, intent, ISO mapping, constraints, acceptance criteria).
\end{itemize}
NL-rich and Structured share identical ISO-grounded content; only representation form differs between them. All three conditions use the same model (\modelid), decoding configuration (temperature~0), HumanEval problem set, and evaluation pipeline. We additionally report a \textbf{Function-Only} (no-NFR) baseline (the bare HumanEval stub with no quality clause) as an NFR-independent context row in Table~\ref{tab:summary}; it lets the reader judge whether the one-line NL-simple baseline already shifts quality relative to specifying no NFR at all. Function-Only is a context reference, not part of the paired RQ1--RQ4 comparisons.

Figure~\ref{fig:design} summarizes the study design at a high level. The \textbf{primary axis} contrasts each intervention against NL-simple (RQ1--RQ3). The \textbf{secondary axis} compares NL-rich vs.\ Structured with ISO content held constant (RQ4). Both axes reuse RobuNFR's ten-variation protocol and the same functional task clause across conditions.

\begin{figure}[t]
  \centering
  \fbox{\parbox{0.92\columnwidth}{\small
  \textbf{Study design (conceptual).}\\
  Functional task (fixed) + NFR block (manipulated).\\
  \textbf{Primary:} NL-rich vs.\ NL-simple; Structured vs.\ NL-simple.\\
  \textbf{Secondary:} Structured vs.\ NL-rich (same ISO object).\\
  Outcomes: Pass@1, ET-Pass@1, density metrics, STDEV across 10 variations.\\
  Model: \modelid, $T{=}0$; benchmark: HumanEval/ET ($n{=}164$).
  }}
  \caption{Experimental factors and comparison hierarchy. Primary comparisons evaluate enrichment; secondary comparison evaluates form under fixed content.}
  \label{fig:design}
\end{figure}

\subsection{Illustrative Prompt Contrast (Code Smell NFR)}
To make the manipulation concrete, Listing~\ref{lst:prompt} sketches the three NFR blocks for one NFR (surface wording varies across the ten prompt templates; ISO fields are identical between NL-rich and Structured). The functional prefix (\emph{``complete the following code:''} plus the HumanEval stub) is identical in all conditions.

\begin{figure}[t]
  \small
  \begin{verbatim}
NL-simple:
  Avoid code smells in the implementation.

NL-rich (excerpt):
  Maintainability (ISO/IEC 25010): modularity and
  analysability. Avoid refactor-worthy patterns; keep
  functions cohesive; limit nesting; ... [constraints
  and acceptance criteria for smell-related quality]

Structured (excerpt):
  {"attribute":"code_smell","iso_25010":{...},
   "constraints":[...],"acceptance_criteria":[...]}
  \end{verbatim}
  \caption{Illustrative NFR blocks for the Code Smell condition (abbreviated). Full templates are in the replication package.}
  \label{lst:prompt}
\end{figure}

The NL-simple phrase is short and leaves interpretation to the model. NL-rich and Structured spell out ISO characteristics, explicit constraints, and acceptance criteria tied to the quality attribute (never to passing HumanEval tests).

\subsection{Holding Content Constant; Varying Only Form}
To make the rich-NL and structured conditions content-equivalent, both are produced from a single source object per NFR (attribute, intent, ISO characteristic/sub-characteristics, constraints, acceptance criteria). The \emph{functional task clause} (\emph{``complete the following code:''}) is byte-for-byte identical across all conditions, so the only manipulated difference is whether the NFR block is prose or JSON. The acceptance criteria are restricted to the quality attribute and never instruct the model to ``pass the tests'', preventing a leakage confound on Pass@1. For each condition we create ten prompt variations (varying the surface realization) to measure prompt sensitivity, mirroring RobuNFR's ten-variation protocol.

\subsection{ISO Mapping}
Each NFR is mapped to ISO/IEC~25010:2023 as follows: \emph{performance} $\rightarrow$ Performance Efficiency (time behavior, resource utilization); \emph{error handling} $\rightarrow$ Reliability (fault tolerance) with relevant exception-handling practice; \emph{code smell} and \emph{readability} $\rightarrow$ Maintainability (modularity/analysability and analysability/modifiability, respectively). The mapping is an explicit design decision documented in the package; we treat its subjectivity as a construct threat (Section~\ref{sec:threats}).

For each NFR, the shared ISO object includes: (i)~a short \emph{intent} statement aligned with the sub-characteristic; (ii)~two to four \emph{constraints} phrased as imperative requirements on generated code; and (iii)~ \emph{acceptance criteria} that remain attribute-specific (e.g., limiting nesting depth for readability) and never reference HumanEval test passage. NL-rich renders these fields as connected prose; Structured emits the same fields as JSON keys. Ten prompt \emph{variations} paraphrase surface wording (synonyms, clause order) while preserving the ISO object, following RobuNFR's sensitivity protocol.

\subsection{Model, Benchmark, and Generation}
We use the model \modelid with temperature~0 (greedy), the only decoding configuration in the study. The NL-simple baseline was collected in an earlier batch within this study (Apr--May 2026); the NL-rich and Structured interventions followed in June 2026. All conditions use the same pinned snapshot and temperature~0. Because batch timing differs, we report separate \emph{post hoc} snapshot-stability checks (Section~\ref{sec:snapshot-stability}) rather than treating the batches as interchangeable without qualification. OpenAI's snapshot policy states that model \emph{snapshots} lock a specific version for consistent behavior~\cite{openai2026gpt5snapshots}; we cite it only as secondary support. Generation targets the HumanEval problem set (164 tasks) via EvalPlus~\cite{liu2023evalplus}. For each (NFR, condition) we generate code for all ten prompt variations, yielding the per-problem completions analyzed below. Generation concurrency is an infrastructure parameter only and does not affect prompts, model, temperature, or any metric.

\subsection{Snapshot Stability Check (Batch Mitigation)}
\label{sec:snapshot-stability}
Because the NL-simple baseline and the June interventions were collected in separate batches, we ran \emph{post hoc} stability checks on NL-simple prompt0 with the same \modelid\ snapshot and pipeline, comparing per-problem Pass@1 against the April--May baseline using paired Wilcoxon tests (10{,}000 bootstrap resamples for mean-difference 95\% CIs; scripts in \wonestabilitydir).

\textbf{Full-set check (Performance, $n=164$).} We re-executed NL-simple prompt0 on all 164 HumanEval tasks in August 2026. Mean Pass@1 was 0.93 (Apr--May) vs.\ 0.92 (August): 97.6\% per-task agreement (one task improved, three worsened), paired Wilcoxon $p=0.424$, Cliff's $\delta=0.01$ (negligible), and bootstrap 95\% CI for the mean Pass@1 difference $[-0.04,+0.01]$, which \emph{includes zero}. ET-Pass@1 was likewise non-significant ($0.84$ vs.\ $0.84$; $p=0.773$). We find no strong evidence of snapshot drift on the complete Performance subset.

\textbf{Pilot subset ($n=30$, June 2026).} Earlier, we re-ran NL-simple prompt0 on a fixed stratified subset (seed~42; pass/fail balanced from the Performance baseline) for Performance and Error Handling in June 2026. On this subset, Performance Pass@1 was 0.63 vs.\ 0.77 ($p=0.072$; bootstrap CI $[+0.03,+0.27]$, excluding zero), while Error Handling was stable ($0.83$ vs.\ $0.87$; $p=1.0$). Because the subset over-samples failing tasks (11 failures exist in the full baseline), its absolute Pass@1 rates are much lower than on all 164 tasks and should not be read as contradicting the full-set August check.

\textbf{Interpretation.} Non-concurrent collection between the April--May baseline and the June interventions remains an internal-validity concern, but the August full-set re-run suggests that snapshot drift on Performance NL-simple prompt0 is unlikely to explain the primary comparisons. We still report primary statistics with their original batch timestamps and interpret the pilot subset as underpowered, exploratory evidence. Concurrent re-collection of the full NL-simple baseline across all NFRs and prompt variations is listed as future work.

\subsection{Metrics}
\label{sec:metrics}
We report: \textbf{Pass@1} and \textbf{ET-Pass@1} (functional correctness on HumanEval and HumanEval-ET); \textbf{exception density}, \textbf{code-smell density}, and \textbf{unreadability density} (issues per ten LOC, from Pylint Refactor/Convention and exception statements); \textbf{LOC}; and \textbf{execution time} (mean over five runs per problem, and the ET variant). To avoid survivorship bias, NFR-quality metrics are compared over a shared set of problems and paired per problem (Section~\ref{sec:stats}).

\subsection{Statistical Analysis}
\label{sec:stats}
The \textbf{primary contrasts} are each intervention (\textbf{NL-rich}, \textbf{Structured}) versus \textbf{NL-simple}, using per-problem paired Wilcoxon signed-rank tests~\cite{wilcoxon1945}, Cliff's delta~\cite{cliff1993} ($\delta>0$ favors the intervention), and Holm--Bonferroni correction~\cite{holm1979} within each (NFR, comparison, metric family). We choose non-parametric paired tests because Pass@1 and density metrics are bounded and often skewed across problems; Wilcoxon respects pairing (same HumanEval task under two prompt conditions) without assuming normality of differences. Cliff's $\delta$ complements $p$-values: several correctness contrasts are statistically significant yet negligible in magnitude (e.g., Error Handling ET-Pass@1), so we report both. Holm--Bonferroni controls family-wise error within each (NFR, comparison, metric family) rather than globally across all tables, matching our pre-specified contrast structure.

The \textbf{secondary contrast} (RQ4) is NL-rich vs.\ Structured with content held constant. Baseline per-problem vectors are loaded from NL-simple evaluation JSONs collected for \modelid; intervention vectors from the June 2026 runs. Per-problem density vectors (unreadability, code-smell, exception) are derived from each JSON's Pylint buckets and exception statements normalized by LOC (Section~\ref{sec:metrics}). \textbf{Robustness} (RQ3): Table~\ref{tab:robustness} reports condition-level STDEV of aggregate metrics across ten prompt variations (descriptive). We additionally apply paired Wilcoxon tests to \emph{per-problem} STDEV (for each task, the sample STDEV of a metric across its ten prompt variations), comparing each intervention to NL-simple within each (NFR, comparison) family of four metrics, with Holm correction (Section~\ref{sec:artifact}). Mean pairwise Jaccard distance remains descriptive; lower diversity on enriched prompts partly reflects shared ISO template boilerplate, not semantic equivalence alone.

\subsection{Reproducibility}
All configurations, prompts, generated code, evaluation outputs, and analysis scripts are released in a replication package (Section~\ref{sec:artifact}). Any execution-time change required to run the pipeline is documented in the package's \texttt{RUNTIME\_NOTES.md}. A traceability map (\resultsnumbers) links each table cell and $p$-value to the originating evaluation JSON or Excel summary, supporting independent verification without re-running generation.

\section{Results}
\label{sec:results}
All numbers below come from the generated result files and are reproduced, with their
sources, in the replication package. We generated and evaluated all three conditions for
\modelid: NL-simple in an earlier batch (Apr--May 2026) and NL-rich and
Structured in June 2026. Every (NFR, condition) cell comprises ten prompt variations over the
164 HumanEval problems. Table~\ref{tab:summary} reports per-condition means; Table~\ref{tab:paired_baseline} the
\textbf{primary} paired tests for correctness and time (each intervention vs.\ NL-simple);
Table~\ref{tab:paired_density} the \textbf{RQ2} paired density tests vs.\ NL-simple;
Table~\ref{tab:paired_form}
the \textbf{secondary} form-only contrast; and Table~\ref{tab:robustness} robustness and
prompt diversity.

\begin{table*}[t]
  \caption{Mean over ten prompt variations per (NFR, condition) on HumanEval. Densities are issues per ten LOC. The Function-Only (no-NFR) row is an NFR-independent context baseline (bare stub, no quality clause). NL-s=NL-simple baseline, NL-r=NL-rich, St=Structured. Source: aggregate Excel files (see \texttt{results\_numbers.json}).}
  \label{tab:summary}
  \small
  \begin{tabular}{llrrrrrrr}
    \toprule
    NFR & Cond. & Pass@1 (\%) & ET-Pass@1 (\%) & Exc.\ density & Smell density & Unread.\ density & LOC & Time (s) \\
    \midrule
    \multicolumn{2}{l}{Function-Only (no NFR)} & 94.5 & 83.4 & 0.02 & 0.06 & 1.85 & 1694.7 & 0.06 \\
    \midrule
    Performance & NL-s & 93.1 & 82.8 & 0.03 & 0.04 & 0.88 & 2736.3 & 0.04 \\
     & NL-r & 94.5 & 82.4 & 0.05 & 0.04 & 0.69 & 3087.4 & 0.08 \\
     & St & 93.4 & 81.8 & 0.05 & 0.04 & 0.79 & 2823.0 & 0.08 \\
    \midrule
    Error Handling & NL-s & 95.1 & 81.0 & 0.99 & 0.05 & 0.57 & 4244.4 & 0.17 \\
     & NL-r & 92.3 & 76.0 & 1.18 & 0.03 & 0.42 & 4692.0 & 0.06 \\
     & St & 93.6 & 78.4 & 1.16 & 0.03 & 0.43 & 4540.8 & 0.06 \\
    \midrule
    Code Smell & NL-s & 96.8 & 84.6 & 0.11 & 0.02 & 0.67 & 2850.5 & 0.05 \\
     & NL-r & 96.3 & 84.2 & 0.13 & 0.01 & 0.53 & 3679.9 & 0.06 \\
     & St & 96.2 & 84.0 & 0.12 & 0.01 & 0.56 & 3360.0 & 0.06 \\
    \midrule
    Readability & NL-s & 97.1 & 84.9 & 0.04 & 0.02 & 0.60 & 3066.5 & 0.06 \\
     & NL-r & 96.8 & 84.8 & 0.02 & 0.02 & 0.55 & 3126.1 & 0.11 \\
     & St & 96.6 & 84.4 & 0.03 & 0.02 & 0.56 & 3091.1 & 0.13 \\
    \bottomrule
  \end{tabular}
\end{table*}

\begin{table*}[t]
  \caption{Primary paired comparison vs.\ NL-simple baseline per problem (Wilcoxon signed-rank, Cliff's $\delta$, Holm-corrected $p$ within each comparison). $\delta>0$ favors the intervention (NL-rich or Structured) over NL-simple.}
  \label{tab:paired_baseline}
  \footnotesize
  \setlength{\tabcolsep}{3.5pt}
  \begin{tabular}{@{}lllrrrr@{}}
    \toprule
    NFR & Comp. & Metric & $n$ & $p$ & $p_{\mathrm{Holm}}$ & $\delta$ \\
    \midrule
    Performance & NL-r vs NL-s & Pass@1 & 164 & 0.263 & 0.527 & 0.046 (neg.) \\
     &  & ET-Pass@1 & 164 & 0.843 & 0.843 & 0.022 (neg.) \\
     &  & Time & 164 & 0.000 & 0.000 & 1.000 (lg.) \\
    \midrule
     & St vs NL-s & Pass@1 & 164 & 0.922 & 0.922 & 0.022 (neg.) \\
     &  & ET-Pass@1 & 164 & 0.182 & 0.364 & 0.002 (neg.) \\
     &  & Time & 164 & 0.000 & 0.000 & 1.000 (lg.) \\
    \midrule
    Error Handling & NL-r vs NL-s & Pass@1 & 164 & 0.025 & 0.025 & 0.004 (neg.) \\
     &  & ET-Pass@1 & 164 & 0.001 & 0.001 & -0.027 (neg.) \\
     &  & Time & 164 & 0.000 & 0.000 & -1.000 (lg.) \\
    \midrule
     & St vs NL-s & Pass@1 & 164 & 0.154 & 0.154 & 0.007 (neg.) \\
     &  & ET-Pass@1 & 164 & 0.024 & 0.048 & -0.004 (neg.) \\
     &  & Time & 164 & 0.000 & 0.000 & -0.999 (lg.) \\
    \midrule
    Code Smell & NL-r vs NL-s & Pass@1 & 164 & 0.536 & 1.000 & 0.033 (neg.) \\
     &  & ET-Pass@1 & 164 & 0.656 & 1.000 & 0.015 (neg.) \\
     &  & Time & 164 & 0.000 & 0.000 & 0.988 (lg.) \\
    \midrule
     & St vs NL-s & Pass@1 & 164 & 0.776 & 1.000 & 0.023 (neg.) \\
     &  & ET-Pass@1 & 164 & 0.600 & 1.000 & 0.010 (neg.) \\
     &  & Time & 164 & 0.000 & 0.000 & 0.988 (lg.) \\
    \midrule
    Readability & NL-r vs NL-s & Pass@1 & 164 & 0.657 & 1.000 & 0.022 (neg.) \\
     &  & ET-Pass@1 & 164 & 1.000 & 1.000 & 0.024 (neg.) \\
     &  & Time & 164 & 0.000 & 0.000 & 0.988 (lg.) \\
    \midrule
     & St vs NL-s & Pass@1 & 164 & 0.898 & 1.000 & 0.029 (neg.) \\
     &  & ET-Pass@1 & 164 & 0.524 & 1.000 & 0.019 (neg.) \\
     &  & Time & 164 & 0.000 & 0.000 & 0.988 (lg.) \\
    \bottomrule
  \end{tabular}
  \\{\scriptsize $\delta>0$: intervention better than NL-simple baseline.}
\end{table*}

\begin{table*}[t]
  \caption{Paired per-problem comparison of quality densities vs.\ NL-simple baseline (Wilcoxon signed-rank, Cliff's $\delta$, Holm-corrected $p$ within each comparison). For density metrics, lower is better: $\delta<0$ favors the intervention.}
  \label{tab:paired_density}
  \footnotesize
  \setlength{\tabcolsep}{3.5pt}
  \begin{tabular}{@{}lllrrrr@{}}
    \toprule
    NFR & Comp. & Metric & $n$ & $p$ & $p_{\mathrm{Holm}}$ & $\delta$ \\
    \midrule
    Performance & NL-r vs NL-s & Unread. & 164 & 0.000 & 0.000 & -0.224 (sm.) \\
     &  & Smell & 164 & 0.893 & 0.893 & -0.021 (neg.) \\
     &  & Exc. & 164 & 0.002 & 0.003 & 0.027 (neg.) \\
    \midrule
     & St vs NL-s & Unread. & 164 & 0.000 & 0.000 & -0.101 (neg.) \\
     &  & Smell & 164 & 0.943 & 0.943 & -0.029 (neg.) \\
     &  & Exc. & 164 & 0.001 & 0.002 & 0.015 (neg.) \\
    \midrule
    Error Handling & NL-r vs NL-s & Unread. & 164 & 0.000 & 0.000 & -0.398 (med.) \\
     &  & Smell & 164 & 0.000 & 0.000 & -0.184 (sm.) \\
     &  & Exc. & 164 & 0.000 & 0.000 & 0.270 (sm.) \\
    \midrule
     & St vs NL-s & Unread. & 164 & 0.000 & 0.000 & -0.354 (med.) \\
     &  & Smell & 164 & 0.000 & 0.000 & -0.179 (sm.) \\
     &  & Exc. & 164 & 0.000 & 0.000 & 0.239 (sm.) \\
    \midrule
    Code Smell & NL-r vs NL-s & Unread. & 164 & 0.000 & 0.000 & -0.343 (med.) \\
     &  & Smell & 164 & 0.049 & 0.099 & -0.048 (neg.) \\
     &  & Exc. & 164 & 0.716 & 0.716 & -0.019 (neg.) \\
    \midrule
     & St vs NL-s & Unread. & 164 & 0.000 & 0.000 & -0.236 (sm.) \\
     &  & Smell & 164 & 0.237 & 0.473 & -0.047 (neg.) \\
     &  & Exc. & 164 & 0.894 & 0.894 & -0.047 (neg.) \\
    \midrule
    Readability & NL-r vs NL-s & Unread. & 164 & 0.000 & 0.000 & -0.163 (sm.) \\
     &  & Smell & 164 & 0.408 & 0.408 & -0.011 (neg.) \\
     &  & Exc. & 164 & 0.000 & 0.001 & -0.055 (neg.) \\
    \midrule
     & St vs NL-s & Unread. & 164 & 0.000 & 0.000 & -0.132 (neg.) \\
     &  & Smell & 164 & 0.587 & 0.587 & -0.022 (neg.) \\
     &  & Exc. & 164 & 0.001 & 0.003 & -0.049 (neg.) \\
    \bottomrule
  \end{tabular}
  \\{\scriptsize $\delta<0$: intervention lower density than NL-simple (improvement for smell/unreadability). For exception density, higher values in the intervention are consistent with the Error Handling NFR's fault-tolerance intent; see Section~5.2 for interpretation.}
\end{table*}

\begin{table*}[t]
  \caption{Secondary paired comparison \emph{Structured vs.\ NL-rich} (form only; content held constant). $\delta>0$ favors Structured.}
  \label{tab:paired_form}
  \footnotesize
  \setlength{\tabcolsep}{3.5pt}
  \begin{tabular}{@{}lllrrrr@{}}
    \toprule
    NFR & Comp. & Metric & $n$ & $p$ & $p_{\mathrm{Holm}}$ & $\delta$ \\
    \midrule
    Performance & St vs NL-r & Pass@1 & 164 & 0.162 & 0.485 & -0.023 (neg.) \\
     &  & ET-Pass@1 & 164 & 0.477 & 0.727 & -0.019 (neg.) \\
     &  & Time & 164 & 0.363 & 0.727 & 0.125 (neg.) \\
    \midrule
    Error Handling & St vs NL-r & Pass@1 & 164 & 0.149 & 0.149 & 0.003 (neg.) \\
     &  & ET-Pass@1 & 164 & 0.040 & 0.080 & 0.021 (neg.) \\
     &  & Time & 164 & 0.001 & 0.002 & 0.164 (sm.) \\
    \midrule
    Code Smell & St vs NL-r & Pass@1 & 164 & 0.458 & 0.915 & -0.011 (neg.) \\
     &  & ET-Pass@1 & 164 & 0.495 & 0.915 & -0.006 (neg.) \\
     &  & Time & 164 & 0.000 & 0.000 & -0.794 (lg.) \\
    \midrule
    Readability & St vs NL-r & Pass@1 & 164 & 0.833 & 1.000 & 0.006 (neg.) \\
     &  & ET-Pass@1 & 164 & 0.551 & 1.000 & -0.005 (neg.) \\
     &  & Time & 164 & 0.000 & 0.000 & 0.964 (lg.) \\
    \bottomrule
  \end{tabular}
  \\{\scriptsize $\delta>0$: Structured better than NL-rich.}
\end{table*}

\begin{table}[t]
  \caption{Robustness (STDEV across ten prompt variations; lower = more stable) and prompt diversity (mean pairwise Jaccard distance). Condition-level STDEV is descriptive; per-problem STDEV inferential tests are in Section~\ref{sec:stats} and the replication package (Section~\ref{sec:artifact}).}
  \label{tab:robustness}
  \small
  \begin{tabular}{llrrrrr}
    \toprule
    NFR & Cond. & Pass@1 & Exc. & Smell & Unread. & Div. \\
    \midrule
    Performance & NL-s & 0.0158 & 0.010 & 0.006 & 0.123 & 0.584 \\
     & NL-r & 0.0057 & 0.006 & 0.007 & 0.037 & 0.123 \\
     & St & 0.0109 & 0.004 & 0.008 & 0.037 & 0.101 \\
    \midrule
    Error Handling & NL-s & 0.0151 & 0.361 & 0.009 & 0.172 & 0.595 \\
     & NL-r & 0.0092 & 0.083 & 0.003 & 0.015 & 0.134 \\
     & St & 0.0087 & 0.100 & 0.004 & 0.017 & 0.109 \\
    \midrule
    Code Smell & NL-s & 0.0076 & 0.042 & 0.008 & 0.096 & 0.437 \\
     & NL-r & 0.0129 & 0.014 & 0.003 & 0.021 & 0.107 \\
     & St & 0.0096 & 0.011 & 0.004 & 0.021 & 0.088 \\
    \midrule
    Readability & NL-s & 0.0050 & 0.013 & 0.009 & 0.102 & 0.549 \\
     & NL-r & 0.0071 & 0.004 & 0.005 & 0.011 & 0.127 \\
     & St & 0.0077 & 0.007 & 0.005 & 0.020 & 0.102 \\
    \bottomrule
  \end{tabular}
\end{table}

\subsection{RQ1: Functional Correctness vs.\ NL-Simple}
Neither intervention reliably \emph{improves} functional correctness over the NL-simple baseline
(Table~\ref{tab:paired_baseline}). For Performance, Code Smell, and Readability, all Pass@1 and
ET-Pass@1 comparisons have negligible Cliff's $\delta$ ($|\delta|\le 0.033$) and are non-significant
after Holm correction. At the aggregate level (Table~\ref{tab:summary}), mean Pass@1 differs by
at most $1.4$ percentage points from NL-simple for any intervention (e.g., Performance NL-rich
$94.5$ vs.\ NL-simple $93.1$).

The exception is \textbf{Error Handling}, where enrichment \emph{hurts} extended-test correctness:
NL-rich ET-Pass@1 drops from $81.0$ to $76.0$ with $p_{Holm}=0.001$ ($\delta=-0.027$, negligible
magnitude but significant); Structured ET-Pass@1 drops to $78.4$ with $p_{Holm}=0.048$. NL-rich
Pass@1 also falls ($95.1\rightarrow 92.3$; raw $p=0.025$, negligible $\delta$). At roughly five
percentage points absolute, the ET-Pass@1 decrease is modest in pass rate but systematic across
problems after pairing. We interpret
this as the richer fault-tolerance specification eliciting exception patterns that conflict with
HumanEval's functional tests on some problems, consistent with RobuNFR's observation that NFRs can
reduce Pass@1~\cite{lin2025robunfr}. HumanEval's oracles emphasize exact input-output behavior on
small functions; they rarely reward defensive try/except structure unless the reference solution
includes it.

\subsection{RQ2: NFR-Specific Code Quality vs.\ NL-Simple}
Where interventions show clear movement is in \textbf{static-analysis quality densities}
(Table~\ref{tab:summary}; paired tests in Table~\ref{tab:paired_density}). \textbf{Unreadability density} falls markedly versus
NL-simple for every NFR in both aggregate means and paired per-problem tests (all eight intervention comparisons have $p_{Holm}<0.05$ with negative $\delta$): Performance $0.88\rightarrow 0.69$ (NL-rich, $-22\%$ relative) and
$0.79$ (Structured, $-10\%$); Error Handling $0.57\rightarrow 0.42$/$0.43$ ($-26\%$/$-25\%$);
Code Smell $0.67\rightarrow 0.53$/$0.56$; Readability $0.60\rightarrow 0.55$/$0.56$. Because
densities are normalized per ten LOC, these shifts are not artifacts of shorter or longer outputs
alone: LOC does increase for some interventions (e.g., Error Handling NL-rich $4244\rightarrow 4692$ mean LOC), which would tend to inflate raw issue counts if unnormalized.

For context, the \textbf{Function-Only} (no-NFR) row in Table~\ref{tab:summary} is \textbf{descriptive context only} (not part of paired RQ1--RQ4): it shows higher unreadability density ($1.85$ per ten LOC) than any NFR condition despite competitive Pass@1 ($94.5\%$). We do not test this contrast inferentially.

For the Code Smell NFR, \textbf{code-smell density}
drops from $0.019$ (NL-simple) to $0.011$ (NL-rich) at aggregate level (42\% relative reduction),
with Structured at $0.012$; the paired per-problem test for NL-rich vs.\ NL-simple is borderline
(raw $p=0.049$, $p_{Holm}=0.10$, negligible $\delta$). Exception density rises for Error Handling
($0.99\rightarrow 1.18$), significant in paired tests ($p_{Holm}<0.05$, positive $\delta$).
We treat this increase as \textbf{ambiguous evidence} rather than a clear quality gain: a higher
exception density is consistent with the NFR's fault-tolerance intent (more explicit input validation
and error paths), but it is also the most likely mechanism behind the ET-Pass@1 drop reported in
Section~5.1, because added \texttt{try}/\texttt{except} and \texttt{raise} statements can conflict
with HumanEval's exact-output oracles. The same signal can therefore indicate either better defensive
design or reduced benchmark compatibility, and the static metric alone cannot disambiguate the two.
For Error Handling, paired tests also show significant code-smell reductions ($p_{Holm}<0.05$).

For \textbf{execution time}, both interventions differ strongly from NL-simple in paired tests
(large $\delta$ in all NFRs), with direction depending on the NFR (e.g., Error Handling
interventions are faster ($\delta\approx -1.0$) while Performance interventions are slower
($\delta=1.0$). Because execution time depends on hardware and is noisy, we report these only as internal
deltas (Section~\ref{sec:threats}).

\subsection{RQ3: Robustness vs.\ NL-Simple}
Table~\ref{tab:robustness} reports condition-level STDEV across ten prompt variations (descriptive). Pass@1 STDEV falls for Performance (NL-simple $0.0158$ to NL-rich $0.0057$; 64\% lower) and Error Handling ($0.0151$ to $0.0092$/$0.0087$), but paired Wilcoxon tests on \emph{per-problem} Pass@1 STDEV do not reach significance after Holm correction (e.g., Performance NL-rich: $p=0.044$, $p_{\mathrm{Holm}}=0.13$).

Unreadability STDEV shows a clearer pattern: aggregate reductions (e.g., Performance $0.123$ to $\approx 0.037$) are confirmed inferentially: all eight intervention--baseline contrasts (four NFRs $\times$ NL-rich and Structured) yield $p_{\mathrm{Holm}}<0.05$ with negative Cliff's $\delta$ on per-problem unreadability STDEV. Several code-smell and exception STDEV contrasts are likewise significant (full inferential results in the replication package; Section~\ref{sec:artifact}). Code Smell Pass@1 STDEV is mixed at the aggregate level (NL-rich slightly higher than NL-simple); Readability Pass@1 STDEV stays low across conditions ($97.1\%$ baseline Pass@1).

Intervention prompts are \emph{shorter and less textually diverse} than NL-simple (Jaccard $0.10$--$0.13$ vs.\ $0.44$--$0.59$), yet exhibit lower unreadability STDEV under both descriptive and inferential summaries; the Jaccard gap partly reflects fixed ISO template text. With RQ2, enrichment appears to steer models toward implementations with similar test outcomes but lower static-quality sensitivity to re-wording.

\subsection{RQ4: Form (Secondary; NL-Rich vs.\ Structured)}
When ISO content is held constant, representation form has \emph{no statistically significant
effect on functional correctness} (Table~\ref{tab:paired_form}): every Pass@1 and ET-Pass@1
comparison has negligible $\delta$ ($|\delta|\le 0.023$) and is non-significant after Holm
correction (smallest $p_{Holm}=0.080$ for Error Handling ET-Pass@1). Quality densities at the
aggregate level differ only slightly between forms (e.g., unreadability Performance $0.69$ vs.\
$0.79$). Form can affect execution time selectively (large $\delta$ for Code Smell and
Readability) but, as above, we treat time cautiously.

\subsection{Synthesis Across Research Questions}
Reading RQ1--RQ4 together yields a coherent pattern rather than four disconnected null results.
\textbf{RQ1} (correctness vs.\ baseline) is largely null except for Error Handling, where richer
fault-tolerance text appears to trade extended-test pass rate for explicit exception structure.
\textbf{RQ2} compares NFR-specific quality against NL-simple (Table~\ref{tab:paired_density}). Unreadability density improves for all four NFRs. Code-smell density improves when the prompt names that attribute, though the paired per-problem test for Code Smell is borderline after Holm correction.
\textbf{RQ3} (robustness vs.\ baseline): condition-level Pass@1 STDEV reductions remain descriptive after Holm correction on per-problem tests, but per-problem unreadability STDEV is significantly lower for enriched prompts in all four NFRs ($p_{\mathrm{Holm}}<0.05$), despite shorter and less lexically diverse prompt text (Table~\ref{tab:robustness}).
\textbf{RQ4} (form) is null for correctness: once ISO content is fixed, JSON and prose are
empirically interchangeable for Pass@1 and ET-Pass@1 on this model and benchmark.

The ``so what'' is therefore not ``Structured beats NL-rich'' or vice versa, but ``ISO-grounded
enrichment changes \emph{quality proxies} and \emph{stability} relative to one-line NFRs without
buying functional correctness, and format is a secondary engineering choice.'' This ordering
matches how practitioners should prioritize effort: invest in standard-grounded \emph{content}
first; choose JSON or prose based on tooling, not expected LLM performance gains.

\section{Discussion}
Our results support a single narrative: \emph{ISO-grounded enrichment improves quality proxies and
robustness relative to terse one-line NFRs, but does not reliably improve functional correctness;
when content is fixed, representation form is secondary.} This section unpacks that claim,
relates it to RobuNFR, and draws implications for researchers and practitioners in component-based
and quality-aware LLM workflows.

\subsection{Enrichment Matters; Correctness Does Not Follow Automatically}
The primary comparison against NL-simple shows a deliberate trade-off rather than uniform failure
or success. ISO-grounded paragraphs and JSON objects add constraints, acceptance criteria, and
explicit mappings to ISO/IEC~25010 characteristics. Those additions give the model a narrower,
more semantically anchored interpretation of the NFR. That anchoring appears to reduce Pylint
Convention issues (unreadability density) and, for the Code Smell NFR, Refactor warnings, without
systematically raising Pass@1 or ET-Pass@1.

For three of four NFRs (Performance, Code Smell, Readability), correctness effects are negligible
($|\delta|\le 0.033$ after Holm). This suggests that enrichment neither ``fixes'' nor ``breaks''
functional synthesis on HumanEval for maintainability- and performance-oriented phrasing, at least
under greedy decoding with a strong contemporary model. Practitioners should not expect
ISO-grounded NFR paragraphs to substitute for tests or to raise benchmark pass rates; the benefit
lies elsewhere (Section~\ref{sec:disc-quality}).

Error Handling is the exception. Richer fault-tolerance specifications significantly reduce
ET-Pass@1 ($p_{Holm}=0.001$ for NL-rich; $0.048$ for Structured) while \emph{increasing}
exception density. We stress that the rise in exception density is \textbf{ambiguous evidence}:
it cannot be read as a clean reliability improvement, because the same added \texttt{try}/\texttt{except}
and validation branches that satisfy the NFR text are the most plausible cause of the ET-Pass@1 drop on
HumanEval's exact-output oracles. In other words, the very signal that looks like NFR satisfaction in a
static view (more explicit error handling) is entangled with the functional regression in a dynamic view.
Disentangling ``better defensive design'' from ``benchmark incompatibility'' would require oracles that
reward fault handling, which HumanEval does not provide (Section~\ref{sec:disc-benchmark-limits}).
This aligns with RobuNFR's finding that NFR-aware prompts can reduce Pass@1~\cite{lin2025robunfr}:
here, the reduction is tied to \emph{more explicit} reliability requirements rather than to
surface re-wording alone. For teams adopting ISO-grounded error-handling NFRs, we recommend
validating against extended tests and domain oracles, not assuming benchmark pass rate as a proxy
for reliability satisfaction.

\subsection{Quality and Robustness Gains}
\label{sec:disc-quality}
Static-analysis densities improve in a direction consistent with the NFR intent. Unreadability
density falls for every NFR (e.g., Performance NL-simple $0.88$ vs.\ NL-rich $0.69$, a 22\%
relative drop at aggregate level). Code-smell density drops when the NFR targets that attribute
($0.019\rightarrow 0.011$ for NL-rich). These are proxy metrics, not user-perceived quality, but
they operationalize maintainability-related NFRs in a way comparable to RobuNFR and prior ISO-oriented
work~\cite{lai2025nfqc}.

Robustness: Table~\ref{tab:robustness} shows lower condition-level Pass@1 STDEV for Performance and Error Handling under enrichment. Per-problem Wilcoxon tests do not confirm Pass@1 STDEV gains after Holm correction, but \emph{do} confirm lower unreadability STDEV for every intervention--baseline pair ($p_{\mathrm{Holm}}<0.05$). Intervention prompts are \emph{shorter} and have \emph{lower} mean pairwise Jaccard distance than NL-simple ($\approx 0.10$--$0.13$ vs.\ $0.44$--$0.59$); part of that gap reflects shared ISO template boilerplate. Lower unreadability STDEV is therefore not explained by more varied prompt text alone; enrichment constrains static-quality outcomes under re-wording even when functional Pass@1 variation remains descriptive.

\subsection{Form Is Secondary When Content Is Fixed}
RQ4 asks whether LLMs ``prefer'' JSON over prose when ISO fields are identical. The answer on
correctness is no: all Pass@1 and ET-Pass@1 contrasts have negligible Cliff's $\delta$ and fail
to reach significance after Holm correction. Aggregate quality densities differ only slightly
(e.g., unreadability Performance $0.69$ vs.\ $0.79$). Teams integrating NFRs from architecture
models, quality portals, or MDE pipelines can serialize requirements as JSON for traceability and
tooling without expecting a correctness penalty relative to the same content in prose \emph{on this
model}. We do not claim universal null effects for all models or tasks; we claim that format was
not the dominant lever in our controlled setting.

\subsection{Per-NFR Patterns}
\textbf{Performance.} Enrichment does not move Pass@1/ET-Pass@1 but lowers unreadability density; Pass@1 STDEV reductions in Table~\ref{tab:robustness} are descriptive after Holm correction on per-problem tests, while unreadability STDEV gains are inferentially supported (Section~5.3). Mean execution time rises vs.\ NL-simple (large positive $\delta$); we treat timing as indicative only (Section~\ref{sec:threats}).

\textbf{Error Handling.} Enrichment increases exception density and reduces ET-Pass@1, while also
lowering unreadability density and prompt sensitivity. The NFR is partially satisfied in static
structure but conflicts with benchmark tests on a subset of problems.

\textbf{Code Smell.} Correctness remains stable; code-smell and unreadability densities improve.
This is the clearest alignment between NFR intent and proxy metric movement.

\textbf{Readability.} Similar to Code Smell for correctness and unreadability; smell density
was already low at baseline, leaving less headroom for improvement.

\subsection{Relation to RobuNFR and the ``One-Line'' Baseline}
Lin et al.~\cite{lin2025robunfr} establish that NFR-aware generation is sensitive to prompt
variation and can reduce Pass@1. Our NL-simple baseline reproduces that sensitivity pattern:
higher Pass@1 STDEV and higher lexical diversity than enriched prompts. RobuNFR does not ask
whether \emph{enriching} the NFR content (beyond re-wording the same terse phrase) changes
quality and robustness; we show that it does for static-analysis proxies and stability, even
when correctness is flat or slightly worse. Comparing interventions only against each other, without
a one-line baseline, would have masked this: RQ4's null form result does \emph{not} imply enrichment
is useless (Section~\ref{sec:disc-implications}).

\subsection{What HumanEval Can and Cannot Tell Us About NFRs}
\label{sec:disc-benchmark-limits}
Our conclusions are bounded by what the benchmark can observe, and this boundary is itself part of
the finding rather than a footnote. HumanEval problems are small, self-contained Python functions with
exact-output unit tests. This design is well suited to \emph{functional} correctness but only weakly
sensitive to most ISO/IEC~25010 quality characteristics. Three consequences follow. First, the NFRs we
can measure here are necessarily the ones with \emph{intra-function, statically observable} signatures
(readability/code-smell via Pylint, local exception structure, execution time of a single call); genuinely
architectural concerns such as modifiability across modules, compatibility, scalability, or security are
out of reach on single-function tasks. Second, the benchmark's oracles actively penalize some legitimate
NFR behavior: a function that validates inputs and raises specific exceptions can be ``more reliable'' in
the ISO sense yet fail an exact-output test, which is exactly the Error Handling tension we observe. The
metric and the oracle thus disagree by construction, not by accident. Third, density proxies are most
meaningful as \emph{relative} within-task contrasts (intervention vs.\ baseline on the same problem),
which is why we pair per problem; they should not be read as absolute maintainability scores.

This framing tempers our claims in a specific way. The RQ2 quality gains and RQ3 stability gains are real
and significant within HumanEval's observable window, but they speak to \emph{function-level} static
quality and prompt sensitivity, not to system-level maintainability or reliability. Likewise, the RQ4 null
form effect holds for short functions; structured specifications might matter more when NFRs constrain
interfaces, configuration, or cross-component contracts that a single-function benchmark cannot express.
We therefore read our results as a lower bound on where representation could matter: if even content-rich
enrichment leaves correctness flat on small functions, the leverage of NFR specification is more likely to
appear in repository- and architecture-scale generation, which we flag as the priority for future
benchmarks (Section~\ref{sec:threats}).

\subsection{Implications}
\label{sec:disc-implications}
\textbf{For researchers.} (i)~Report enrichment level explicitly when studying NFR-aware LLM
generation; a null form-only comparison is insufficient to characterize ISO-grounded interventions.
(ii)~Pair functional benchmarks with NFR-aligned proxy metrics and extended tests; Error Handling
shows they can diverge. (iii)~Measure prompt sensitivity alongside correctness; robustness is an
outcome variable in its own right for NFR studies.

\textbf{For practitioners.} (i)~Adopt ISO/IEC~25010-grounded NFR text when the goal is cleaner,
more stable generated code along maintainability dimensions, not higher HumanEval pass rates.
(ii)~Choose JSON vs.\ prose based on pipeline integration, not expected LLM gains. (iii)~For
reliability NFRs, review generated exception structure against project oracles; richer specs may
reduce benchmark compatibility while increasing explicit fault-handling code. (iv)~\textbf{Cost of specification:} ISO objects require upfront authoring (intent, constraints, acceptance criteria mapped to \isoqm). That effort pays off when teams already maintain quality models, architecture portals, or MDE pipelines that export structured attributes; for one-off scripts, a one-line NFR may suffice unless static-quality stability matters (an engineering-effort trade-off, not an API-cost claim).

\textbf{For CBSoft/SBCARS reuse contexts.} Component generators and architecture-centric workflows
often maintain structured quality attributes alongside functional interfaces. Our results suggest
exporting those attributes into LLM prompts is worthwhile for quality and stability, provided
teams validate functional behavior independently.

\subsection{Practical Decision Guide}
Table~\ref{tab:summary} and the paired tests support a simple decision guide. If the goal is
\emph{maximize HumanEval pass rate}, NL-simple and enriched prompts perform similarly for three NFRs,
and enriched Error Handling may hurt ET-Pass@1; a one-line NFR is not clearly worse and may avoid
spec-test conflicts. If the goal is \emph{reduce Pylint Convention/Refactor issues} or \emph{stabilize
outputs across re-wordings}, ISO-grounded enrichment is preferable regardless of JSON vs.\ prose.
If the goal is \emph{tool integration}, Structured JSON is supported without correctness penalty
relative to NL-rich in RQ4. These recommendations are conditional on \modelid and
HumanEval; they illustrate how the primary/secondary RQ ordering translates into engineering choices
rather than treating format comparison as the paper's main contribution.

\section{Threats to Validity}
\label{sec:threats}
\textbf{Construct validity.} NFR-quality is operationalized through static-analysis proxies (Pylint Refactor/Convention densities, exception statements) and execution time rather than human judgment. These proxies align with RobuNFR~\cite{lin2025robunfr} and ISO-oriented LLM quality work~\cite{lai2025nfqc} but do not capture runtime reliability, security, or user-perceived maintainability. \textbf{Exception-statement density} is an ambiguous proxy for reliability: models may add broad or low-quality \texttt{try}/\texttt{except} blocks without improving fault tolerance (Section~5.2). Mean pairwise Jaccard distance mixes surface variation with fixed ISO template text; lower diversity on enriched prompts should not be read as semantic equivalence alone. The NFR$\rightarrow$ISO/IEC~25010 mapping is a researcher decision; alternative mappings (e.g., placing error handling under Maintainability) could change prompt emphasis. We make the mapping explicit and traceable in the replication package. Acceptance criteria never instruct the model to pass tests, and the functional task clause is identical across conditions. For error handling, Pass@1 can penalize spec-conflicting exceptions, so we additionally rely on exception density and ET-Pass@1 and discuss the trade-off in Section~6.

\textbf{Internal validity.} Code-quality metrics are normalized per ten LOC; correctness comparisons are paired per problem. Decoding is greedy (temperature~0). A central threat is \textbf{non-concurrent collection}: the NL-simple baseline predates NL-rich and Structured by several weeks (Apr--May vs.\ June 2026). We fixed the same model snapshot (\modelid), temperature~0, benchmark, and evaluation pipeline across conditions to limit API drift, and OpenAI documents that snapshots lock model versions~\cite{openai2026gpt5snapshots}. We re-ran NL-simple prompt0 as a stability control (Section~\ref{sec:snapshot-stability}): on all 164 Performance tasks (August 2026), Pass@1 was 0.93 vs.\ 0.92 ($p=0.424$; bootstrap 95\% CI $[-0.04,+0.01]$, including zero); a prior June pilot on 30 stratified tasks showed a larger apparent shift that we attribute to subset bias ($n=30$, Performance $p=0.072$). We therefore treat snapshot drift on Performance as unlikely to dominate the primary comparisons, while noting that the baseline and interventions were still not collected concurrently and that Error Handling was assessed only on the small pilot. Residual batch effects unrelated to the model snapshot (e.g., infrastructure changes) cannot be ruled out; the replication package documents pipeline versions, hosts all raw outputs, and includes W1 stability JSONs (\texttt{\seqsplit{w1\_stability\_comparison.json}}, \texttt{\seqsplit{w1b\_stability\_performance\_164.json}}) in \wonestabilitydir.

\textbf{Conclusion validity.} We use non-parametric paired Wilcoxon signed-rank tests~\cite{wilcoxon1945}, report effect sizes (Cliff's delta), and correct for multiple comparisons (Holm--Bonferroni) separately within each (NFR, comparison, metric family). Correctness, time, and density metrics all use per-problem paired vectors ($n=164$). For density metrics where lower is better, $\delta<0$ favors the intervention. Significance with negligible $\delta$ (Error Handling ET-Pass@1; Code Smell paired smell density) is interpreted as a small but systematic shift across many problems, not a large practical swing.

\textbf{External validity.} One model (\modelid), one benchmark (HumanEval/ET), four NFRs; findings are tied to this snapshot and may not generalize to repository-level generation, multi-file tasks, or weaker models. As discussed in Section~\ref{sec:disc-benchmark-limits}, HumanEval's small single-function tasks and exact-output oracles make it only weakly sensitive to system-level ISO/IEC~25010 characteristics, so our quality and form findings are bounded to function-level static quality and prompt sensitivity. We position the work as extending RobuNFR's prompt-variation methodology with ISO-grounded enrichment and a secondary form factor; MBPP and HumanEval-NFR replication, and repository-/architecture-scale NFR benchmarks, are listed as future work.

\textbf{Reliability/Reproducibility.} Execution time depends on hardware; we compare only internal deltas within the same experimental run. The full pipeline and data are released in the replication package (Section~\ref{sec:artifact}), including \resultsnumbers\ traceability from tables to source files.

\section{Conclusion and Future Work}
We compared two ISO/IEC~25010:2023-grounded NFR interventions (NL-rich prose and Structured JSON)
against the RobuNFR-style NL-simple baseline for four NFRs on HumanEval/ET with \modelid.
\textbf{Primary result:} ISO-grounded enrichment improves static-analysis quality densities and reduces prompt
sensitivity versus one-line NFRs, but does not reliably improve functional correctness and, for Error Handling, may
reduce extended-test pass rate. \textbf{Secondary result:} when content is held constant, prose
versus JSON form has negligible effect on correctness. In one sentence: \emph{what} you specify in
an NFR (standard-grounded enrichment) matters more for quality and robustness than \emph{how} you
serialize it (NL-rich vs.\ JSON), while functional correctness remains dominated by the functional
task and benchmark oracles.

Future work includes MBPP replication, additional models and decoding settings beyond \modelid, human evaluation of ISO-aligned quality, and direct measurement of
requirement ambiguity in NL-simple vs.\ enriched prompts. Extending the study to architecture-centric
benchmarks (e.g., HumanEval-NFR~\cite{han2024archcode}) would test whether enrichment benefits
transfer when functional and non-functional requirements are co-specified.

\section*{Artifact Availability}
\label{sec:artifact}
A replication package (prompt configurations, generated completions, evaluation outputs, analysis scripts, documentation; MIT License) is available at

\noindent\hspace*{-2pt}{\small\url{\artifacturl}}\hspace*{-2pt}.

\begingroup\sloppy
The package includes \resultsnumbers\ (table and $p$-value traceability) and \rqthreestdev\ (RQ3 per-problem STDEV inferential tests).
W1 snapshot-stability checks are in \texttt{\seqsplit{w1\_stability\_comparison.json}} (pilot, $n=30$) and \texttt{\seqsplit{w1b\_stability\_performance\_164.json}} (Performance, $n=164$), under \wonestabilitydir.
\endgroup

\section*{Acknowledgements}
For partially supporting this work, we would like to thank INES.IA (National Institute of Science and Technology for Software Engineering Based on and for Artificial Intelligence) www.ines.org.br, CNPq grant 408817/2024-0, and CAPES (Coordena\c{c}\~{a}o de Aperfei\c{c}oamento de Pessoal de N\'{i}vel Superior, Brazil) for master's scholarship support under the Demanda Social program.

\appendix
\section{ISO Prompt Object Schema (Replication)}
\label{app:schema}
Each NL-rich and Structured prompt is generated from one JSON object per (NFR, variation). Keys
are identical across forms; only serialization differs. Core fields:
\begin{itemize}
  \item \texttt{attribute}: NFR identifier (\texttt{performance}, \texttt{error\_handling}, \texttt{code\_smell}, \texttt{readability}).
  \item \texttt{intent}: one-sentence goal aligned with ISO/IEC~25010 sub-characteristics.
  \item \texttt{iso\_25010}: nested \texttt{characteristic} and \texttt{sub\_characteristics} strings.
  \item \texttt{constraints}: list of imperative rules (typically three or four) scoped to the quality attribute.
  \item \texttt{acceptance\_criteria}: list of checkable conditions; never reference HumanEval tests.
\end{itemize}
Variations alter lexical choice and clause order in \texttt{intent} and constraints while keeping \texttt{iso\_25010} and acceptance criteria fixed. NL-simple prompts use RobuNFR-style one-line templates in \texttt{approach/} (\texttt{nfr\_prompts.py}).

\clearpage
\bibliographystyle{ACM-Reference-Format}
\bibliography{references}

\end{document}